\documentclass[structabstract]{aa}
\usepackage{graphicx}
\usepackage{amssymb}
\usepackage{amsfonts}
\usepackage{txfonts}
\usepackage{natbib}
\begin{document}
\title{The impact of binary stars on the colors of high-redshift galaxies}
%\subtitle{High-redshift galaxies with binary stars.}

\author{Y. Zhang\inst{1}
% \thanks{e-mail:zhy@xao.ac.cn}
  \and   J. Liu\inst{1}
  \and   F. Zhang\inst{2,3}
  \and   Z. Han\inst{2,3}
         }
 \institute{National Astronomical Observatory/Xinjiang Observatory, the Chinese Academy of Sciences, Urumqi 830011, China\\email: {zhy@xao.ac.cn}
 \and National Astronomical Observatory/Yunnan Observatory, the Chinese Academy of
 Sciences, Kunming 650011, China
 \and Key Laboratory of the Structure and Evolution of Celestial Objects, Chinese Academy of Sciences, Kunming 650011, China
 }
 \date{Received ; accepted}
 %******************************************************************************
 % \abstract{}{}{}{}{}
% 5 {} token are mandatory
\abstract
{Evolutionary population synthesis (EPS) models play an important role in many studies on the formation and evolution of galaxies. Most EPS models are still poorly calibrated for certain stellar evolution stages, especially for the treatment of binary stars, which are very different from single stars.}
{We aim to present color-magnitude (C-M) and color-color (C-C) relations for passive model galaxies in the redshift z$\sim0.0-3.0$ and to study the effect of binary interactions on these relations for high-redshift passive galaxies.}
 {Assuming exponentially declining star formation rate, we used a set of theoretical galaxy templates obtained from Yunnan EPS models (with and without binary interactions) to present the C-M and C-C relations for passive galaxies via Monte Carlo simulation. In Yunnan EPS models with binary interactions, various processes are included, such as mass transfer, mass accretion, common-envelope evolution, collisions, supernova kicks, tidal evolution, and all angular momentum loss mechanisms. In these models, approximately 50 per cent of the stellar systems are binary systems with orbital periods less than 100\,yr. This fraction is a typical value for the Milky Way.}
{We find that the inclusion of binary interactions in the model galaxies' spectra can dramatically alter the predicted C-M and C-C relations and their evolution with redshift. For $z\sim 0.0$ and $1.0$, the binary interactions have a minor effect on the C-M and C-C relations, but at $z\sim 2.0$ and $3.0$ the binary interactions have a major effect on the C-M and C-C relations. Especially for the redshift $z\sim 2.0$, the $g-$band magnitude becomes smaller by $1.5$\,mag, the $g-r$ color becomes bluer by $1.0$\,mag, and the $u-g$ color becomes redder by $1.0$\,mag when binary interactions are included.}
{}
%**************************************************************************************************

\titlerunning{High-redshift galaxies with binary stars}
\authorrunning{Y. Zhang et al.}
\maketitle
%
%________________________________________________________________

\section{Introduction}
\label{sec_intro}
Understanding the formation and evolution of galaxies is one of the important and
challenging questions of cosmology. According to the standard cold dark matter (CDM) paradigm, galaxies are initially formed in the center of small CDM halos via gas cooling and subsequent star formation, gradually assembled with time through hierarchical processes, and then evolved into populations with various size, color, and morphology \citep{whi78}. During the cosmic time, the era of $1 < z < 3$ is a crucial stage in terms of star formation, stellar mass content, and galaxy morphology. In this period, the star-forming activity in the Universe and the bulk of stellar mass assembly in galaxies are at their peak levels \citep{dic03,hop06,arn07}. Meanwhile, a variety of observations suggest that the cosmic star formation rate (SFR) density is at its maximum value \citep{dad05,ric06,arn07} at z$\sim2.0$. A number of deep and wide-field multiwavelength surveys have been employed in the past several years to assemble multiwavelength observations of high-redshift galaxies. Therefore,
the star formation history, age, and stellar mass of high-redshift galaxies are well studied through fitting observed  multiband spectral energy distributions (SEDs) with evolutionary population synthesis (EPS) models.

Evolutionary population synthesis is one of the techniques to study galaxy evolution at all eras \citep[][hereafter BC03]{tin78,fio97,vaz99,zha04,mar05,mar07,con09,bru03}.
These EPS models play an important role in deriving the star formation histories, stellar properties, and redshifts from photometry and spectra. So our understanding of stellar population properties, galaxy growth across the cosmic time \citep{sha05,van10}, and the evolution
of stellar mass density \citep{mar09,gon10} is heavily dependent on EPS models.

Most EPS models neglect the effect of binary interactions on some stellar evolution stages. Meanwhile, observations show that binary stars are very common in nearby star clusters and galaxies \citep[][]{abt83,kro11}. For example, \citet{rag10} presented the results of a comprehensive evaluation of the multiplicity of solar-type stars in the solar neighborhood, and their analysis showed that the binary fraction of the progenitor population was about ($50 \pm 4$)\%. \citet{sol07} analyzed the binary population of $13$ low-density Galactic globular clusters with the aim of studying their frequency and distribution. Their study revealed that these globular clusters hold a fraction of binaries ranging from $10$ to $50$ per cent depending on the cluster. \citet{sol10} used deep wide-field photometric observations to derive the fraction of binary systems in a sample of five high-latitude Galactic open clusters. They found that the estimated global fractions of binary systems ranged from $35$ to $70$ per cent depending on the cluster. \citet{min13} analyzed binary fractions and binary distributions of dwarf spheroidal galaxies. They found that binary fractions of Fornax, Sculptor, and Sextans dSphs were consistent with that of Milky Way (about $50$\%) within $63$\% confidence limits.
When binary stars are considered in EPS models, they can enhance the ultraviolet (UV) passbands by $2.0-3.0$ mag for stellar populations at an age of about $1.0$\,Gyr \citep{zha04,zha05}. Moreover, \citet{han07} concluded that most of UV sources in elliptical galaxies came from binary channels. Therefore, neglecting binaries in EPS models can lead to an underestimation of the SED in UV passbands and then affect the determination of parameters for stellar population systems \citep{zfh12,zhy12}.

According to the CDM model cosmology calculator of \citet{wri06}, at redshift z$\sim2.0$ the age of the Universe is about 3.0\,Gyr, and passive galaxies at this redshift should contain young stellar populations ($< 3.0$\,Gyr). In addition, there exist passive galaxies at redshift z$\sim2.0$ \citep{dad04,kri09,mar10,fang12} that form a red sequence at z$\sim2.3$ \citep{kri08}. The spectral shapes of passive galaxies at z$\sim2.0$ indicate that most of their stars have formed over a short timescale in an intense starburst, which verifies that these galaxies are in a post-starburst phase \citep{kri08}. The future evolution of passive galaxies at z$\sim2$ is unknown. Meanwhile, \citet{zha10} showed that the binary interactions have significant influence on the far-UV ($F\rm_{UV}$) band for stellar populations at the age range of $0.5-3.0$\,Gyr. The stellar populations in this age range are just starting to exist in passive galaxies at z$\sim2.0$, hence the effect of binary interactions is obvious for the observed-frame optical passbands when the rest-frame UV emission is shifted to optical wavelength. The evolution of the age of populations, the trend of SFR, the chemical enrichment, and morphology can bring changes in the spectra, as well as the changes in galaxy luminosity and colors. Meanwhile, most of the galaxy properties are correlated with EPS models, and the binary interactions in EPS models can affect the stellar populations. In this study we will discuss the effect of binary interactions on the SED of passive galaxies.

To show the quantitative influence of binary interactions on
the predicted galaxy magnitudes and on the colors at different redshifts,
we use the galaxy template spectra based on Yunnan EPS models, the reddening law, the filter set, and the cosmological parameter through a Monte Carlo (MC) simulation to produce the passive model galaxies, and to study the evolution of color-magnitude (C-M) and color-color (C-C) relations of passive galaxies with redshift.

The structure of this paper is as follows. In Sect. \ref{sect:method}, we briefly describe the method used to generate galaxy sample with different redshift. In Sect. \ref{sect:results}, we show the C-M
and C-C relations predicted by EPS models for different redshifts, followed by discussion. The conclusion is given in Sect. \ref{sect:conclusions}.
\section{Method}
\label{sect:method}
We used the relevent parameters based on the MC simulation to produce the model galaxies. The relevant parameters in this procedure are: (i) the set of galaxy template spectra, including the SED with SFR and  different ages; (ii) the reddening law, which is implemented to account for the effect of interstellar dust on the shape of the SED; (iii) the filter set; and (iv) the standard dark matter model cosmological parameters $H_{0}$, $\Omega_{M}$, and $\Omega_{\Lambda}$. In this section, we will describe these relevant parameters.

\subsection{The stellar population models}
Evolutionary population synthesis is one of the techniques of modeling the spectroscopic and photometric
properties of stellar populations using the knowledge of stellar evolution.
This technique was first introduced by \citet{tin68} and has been developed
rapidly ever since. Moreover, EPS models can be used to build galaxy template
spectra. Recently, binary interactions have also been incorporated in
EPS models by the Yunnan group \citep[Yunnan EPS models;][]{zha04,zha05,zha06}.
To quantify the effect of binary interactions on the predicted
galaxy magnitudes and colors at different redshifts,
we build a theoretical galaxy template SED using
Yunnan models of single stellar populations
\citep[Model A,][without binary interactions]{zha04}
and the models of binary stellar populations
\citep[Model B,][with binary interactions]{zha05}.
These models present the SEDs of stellar populations with and without binary interactions at 90 ages, and the ages are in the range from log($t_{i}\rm /yr$)$=5.000$ to $10.175$.

The Yunnan EPS models were built on the basis of the Cambridge stellar evolution tracks \citep{eggleton71,eggleton72,eggleton73}, BaSeL-2.0 stellar atmosphere models \citep{lejeune97,lejeune98}, and various initial distributions of stars. The Cambridge stellar evolution tracks are obtained by the rapid single/binary evolution codes \citep{hur00,hur02}, which is based on the stellar evolutionary track by \citet{pol98}. In the binary evolution code, various processes are included, such as mass transfer, mass accretion, common-envelope evolution, collisions, supernova kicks, tidal evolution, and all angular momentum loss mechanisms.The main input parameters of the standard models are as follows:\\
(1) The IMF of the primaries gives the relative number of the primaries in the mass range $M \rightarrow M +$ d$M$. The initial primary-mass $M_1$ is given by
\begin{equation}
M_1 = \frac{0.19X}{(1-X)^{0.75} + 0.032 (1-X)^{0.25}},
\label{eq:imfms79-app}
\end{equation}
where $X$ is a random variable uniformly distributed in the range [0, 1].
The distribution is chosen from the approximation to the IMF of \citet{miller79} as given by \citet{eggleton89}
\begin{equation}
\phi(M)_{_{\rm MS79}} \propto \Biggl\{ \matrix{
          M^{-1.4}, & 0.10 \le M \le 1.00 \cr
          M^{-2.5}, & 1.00 \le M \le 10.0 \cr
          M^{-3.3}, & 10.0 \le M \le 100 \cr
          }
\label{eq:imfms79}
\end{equation}
in which $M$ is the stellar mass in units of M$_{\rm \odot}$.\\
(2) The initial secondary-mass distribution, which is assumed to be correlated with the initial primary-mass distribution, satisfies a uniform distribution
\begin{equation}
n(q)=1.0,\,\,\, 0.0\le q \le 1.0,
\label{eq:nq}
\end{equation}
where $q=M_{2}/M_{1}$.
\\
(3) The distribution of orbital separation (or period) is taken as constant in log$a$ (where $a$ is the separation) for wide binaries and fall off smoothly at close separations
\begin{equation}
a{\rm n}(a) = \left\{ \begin{array}{ll}
                           a{\rm_{sep}} (a/a_{0})^{m}, \, a \leq a_{0}\\
                  a{\rm_{sep}}, \,\,\,\, a_{0} < a <a_{1}
                 \end{array}
               \right.
\label{eq:disa}
\end{equation}
in which $a_{\rm sep} \sim 0.070, a_0 = 10 {\rm R_\odot}, a_1 = 5.75 \times 10^6 {\rm R_\odot}$  $=0.13$pc, and $m \sim 1.2$
\citep[][]{han95}.\\
(4) The eccentricity distribution satisfies a uniform form $e\,=\,X$, $X\in $[0, 1].\\

Some of the relevant features of these models can be found in \citet{zha04}
for Model A and \citet{zha05} for Model B. In Model B, 50 per cent of the stars in each stellar population are in binary systems
with orbital
periods less than 100yr. And this fraction is the typical value
for the Galaxy.
We note that both of the two EPS models have the same star sample
($2.5 \times 10^{7}$ binary systems). We assume that all stars are born
at the same time.

\subsection{Theoretical galaxy template}
\label{ssect:template}
The galaxy template should not only comprise the SEDs of different ages, but should also include the star formation history (SFH) of a galaxy. The EPS models only provide the SEDs of stellar populations without any SFR at different ages. Therefore, at a given age we need to generate the SEDs of galaxies with different SFHs by means of EPS models. Several studies have found that the observational properties of local field galaxies with different SFHs can be roughly matched by a population with different SFRs. For example, \citet{ken86} used an exponentially declining SFR to describe the local spirals, and their results could explain the
observations well. In this work, the SFH of passive galaxies is described by a widely used \citep{bru83,pap01,sha05,lee09,wuy09} exponentially declining SFR
\begin{equation}
\label{eq:e-decling-sfr}
\psi(t) \propto exp(-t/\tau) ,
%\psi(t) = \tau^{-1} exp(-t/\tau) ,
\end{equation}
where $\tau$ and $t$ are the $e-$folding time scale and the age of population, respectively. We focus on the passive galaxies, and the range of $\tau$ is from $0.01$\,Gyr to 1.0\,Gyr: $\tau = 0.01, 0.05, 0.1, 0.2,\\ 0.3, 0.4, 0.5, 0.6, 0.7, 0.8, 0.9,$ and $1.0$\,Gyr. The BC03 software package provides the code to construct the galaxy template with different SFRs. Using the stellar population models of Yunnan EPS models and the $e-$folding SFR, we use the BC03 package to build the galaxy templates. For Model A and Model B we also assume that $z=0.02$, and the age range is from 10$^{5}$ yr to 10$^{10.175}$ yr.

\subsection{Definitions of other parameters}
Other parameters used in this work are given as follows:
\begin{itemize}
\item
The redden law of \citet{cal00} is used. The observed shape of the stellar emission $F\rm_{o}(\lambda)$ is reddened using the starburst reddening curve $k(\lambda)=A(\lambda)/E_{s}(B-V)$ with the  standard formulation \citep{cal94}
\begin{equation}
\label{eq:redden-raw}
F_{o}(\lambda)=F_{i}(\lambda)10^{-0.4E_{s}(B-V)k(\lambda)} ,
\end{equation}
where $F_{o}(\lambda)$ and $F_{i}(\lambda)$ are the observed and intrinsic stellar continuum flux densities, respectively. The color excess of stellar continuum $E_{s}(B-V)$ is linked to the color excess ($E(B-V)$) derived from the nebular gas emission lines $E(B-V)$ via $E_{s}(B-V)=(0.44\pm0.03)E(B-V)$ \citep{cal97}. The expression of $k(\lambda)$ is
\begin{equation}
\label{eq:k-lambda}
k(\lambda)= \left\{ \begin{array}{ll}
                   2.695 \left( -2.156 + \frac{1.509}{\lambda} -
                   \frac{0.198}{\lambda ^{2}} + \frac{0.011}{\lambda^{3}}
                   \right) + R_{V}, \\ \;\;\;\;\;\;\;\;\;\;\;\;\;\;\;\;\;\;\;\;\;\;\;\;\;\;(0.12 \mu m \leq \lambda \leq 0.63 \mu
                   m)\\
             2.659 \left( -1.857 + \frac{1.040}{\lambda} \right) + R_{V}, \\\;\;\;\;\;\;\;\;\;\;\;\;\;\;\;\;\;\;\;\;\;\;\;\;\;\;(0.63
             \mu m \leq \lambda \leq 2.20 \mu m)
            \end{array}
        \right. .
\end{equation}
In this work, the extinction $A_{V}$ is allowed to vary from $0$ to $3$ in steps of $0.2$, which corresponds to $E(B-V)$ varying from $0$ to $0.74$ according to the reddening law ($R\rm_{V} = 4.05$) of \citet{cal00}.
\item
We have selected the Sloan Digital Sky Survey (SDSS) standard filter system $u g r i z$ with air mass of $1.3$. Table \ref{tbl:filter} presents the characteristics of these filters, the effective wavelength $\rm \lambda_{eff}$, and the full width at half maximum (FWHM).

\begin{table}[!htp]
\begin{center}
\caption{Properties of SDSS filters: the effective wavelength $\rm \lambda_{eff}$ and FWHM.\label{tbl:filter}}
\begin{tabular}{ccc}
\hline\hline
Filters & $\rm \lambda_{eff}$ ($\rm \AA$) & FWHM ($\rm \AA$) \\
\hline
$u$ &3551 &581 \\
$g$ &4686 &1262\\
$r$ &6166 &1149\\
$i$ &7480 &1237\\
$z$ &8932 &994 \\
\hline
\end{tabular}
\end{center}
\end{table}
\item
Finally, we have adopted a set of standard dark matter model cosmology parameters ($\Omega\rm_{\Lambda}, \Omega\rm_{M}, H_{0}$) $= 0.7, 0.3, 70.0$.
\end{itemize}

\section{Results and discussion}
\label{sect:results}
By means of the MC simulation and some relevant parameters, we produced the passive model galaxies with the $e-$folding SFR at different redshifts. In this section, we give the C-M and C-C relations of passive model galaxies at different redshifts and investigate the effect of binary interactions on the predicted magnitudes and colors. Considering the influence of binary interactions on the observed-frame optical passbands for $z\sim2$, we only focus on the optical passbands. We note that the $ugriz$ Vega magnitudes are in observed-frame.

\subsection{Binary interactions on magnitudes and colors}
The model galaxies at $z\sim 0.0, 1.0, 2.0$, and 3.0 correspond to the galaxies at the range of $0.0\leq z \leq 0.3$, $0.7\leq z \leq 1.3$, $1.7\leq z \leq 2.3$, and $2.7\leq z \leq 3.0$, respectively. We take the range of $2.7\leq z \leq 3.0$ as $z\sim 3.0$, because the colors always lie outside the $u-$band for high redshift galaxies \citep[$z\sim 3.4$,][]{lee09}. In this work, we generate 10 000 model galaxies for each redshift with each model (Model A and Model B). We assume that the formation redshift of galaxies is at z$=10.0$, corresponding to about 0.48\,Gyr for the Universe.

\begin{figure*}[!htp]
\begin{center}
   \includegraphics[bb=30 25 800 700,height=12.cm,width=12.cm,clip,angle=0,scale=0.5,angle=0]{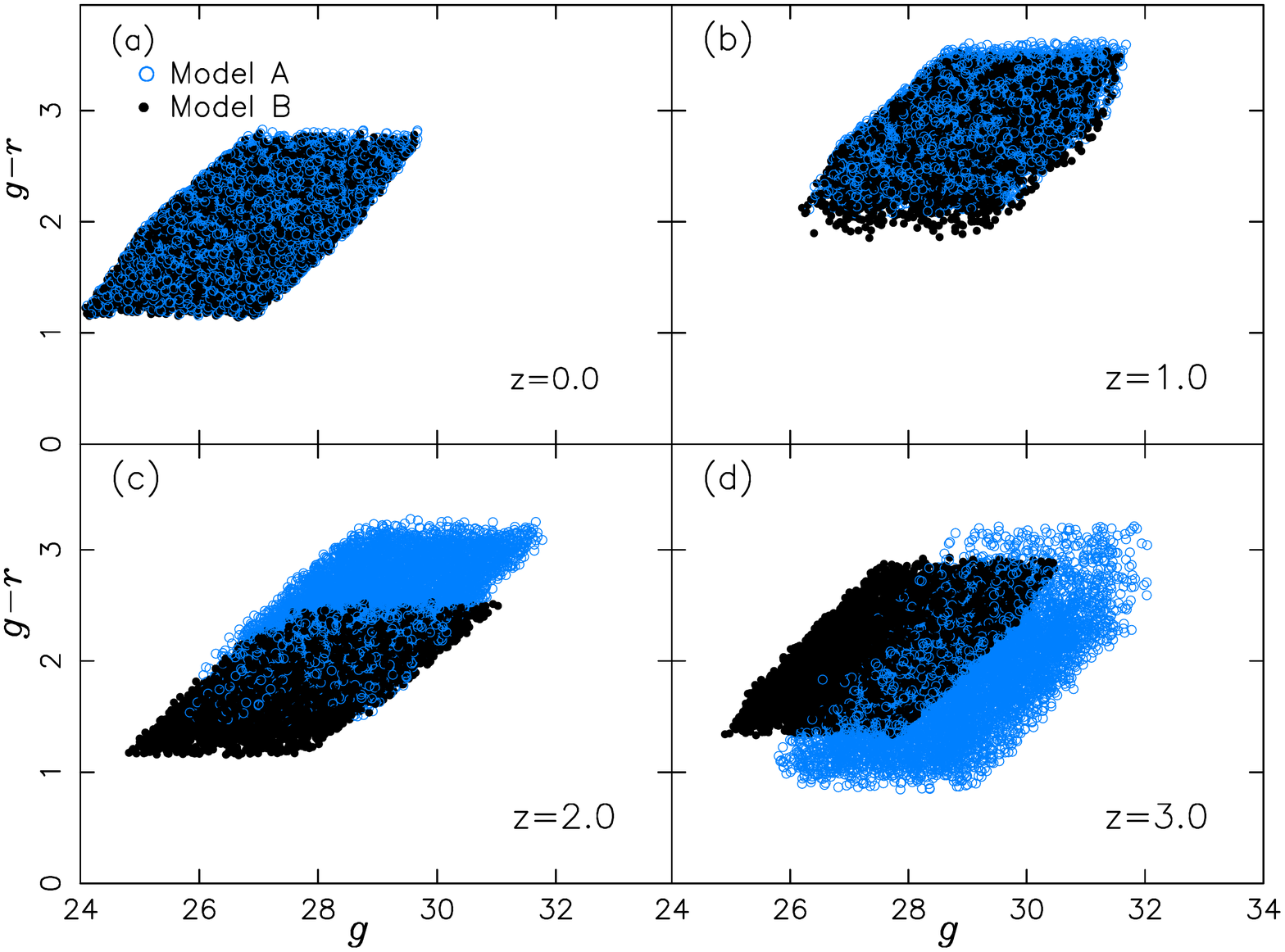}
   \includegraphics[bb=15 25 800 700,height=12.cm,width=12.cm,clip,angle=0,scale=0.5,angle=0]{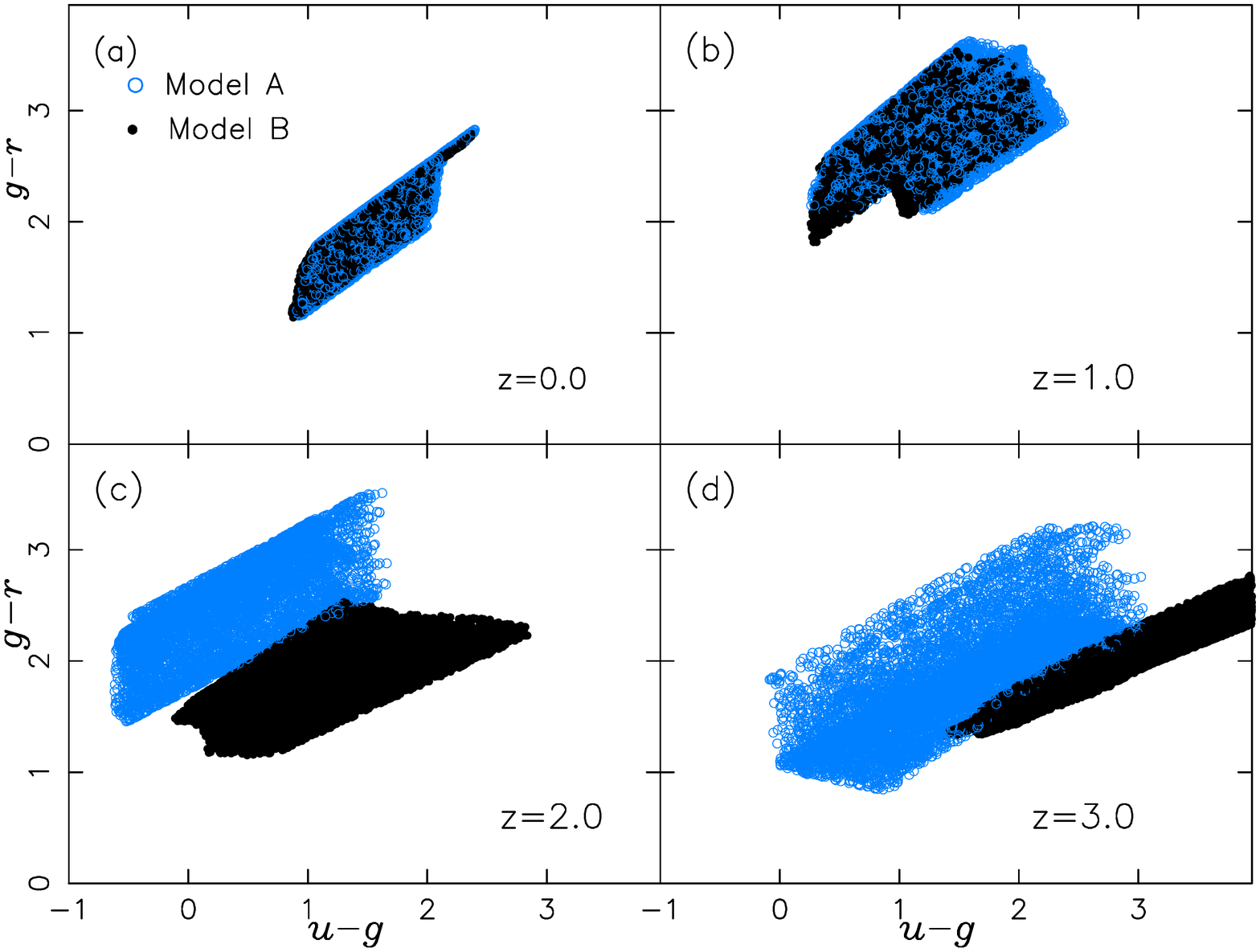}
   \caption{The C-M ($g-r$ color versus $g$ magnitude) and C-C ($g-r$ color versus $u-g$ color) relations for passive galaxies at redshifts of $0.0, 1.0, 2.0$, and $3.0$. Left- and right-hand panels show the C-M and C-C relations, respectively. The blue open circles and black solid circles represent the galaxies generated by Model A and Model B, respectively.}
   \label{fig:a-subsample}
\end{center}
\end{figure*}

Figure \ref{fig:a-subsample} shows the observed-frame $g-r$ color versus $g$ magnitude and $u-g$ color for passive galaxies at four different redshifts ($z=0.0, 1.0, 2.0$, and $3.0$) based on Model A and Model B. The blue open and black solid circles represent the model galaxies based on Model A and Model B, respectively. The left- and right-hand show the C-M and C-C relations, respectively. We can see that the model galaxies based on Model A and Model B
overlap for both the C-M and C-C relations
at $z\sim 0.0$ [panels (a)] and $1.0$ [panels (b)].
%However, there exist some impressive offset ($\Delta$mag$=|$mag$\rm_{B}$-mag$\rm_{A}|$) %for the model galaxies between Model A
%and Model B at $z\sim 2.0$ and $3.0$.
However, it shows a
significant offset for both the C-M and C-C relations for those using Model A and Model B at $z\sim 2.0$ [panels (c), especially for
the C-C relation]. The offset at $z\sim 3.0$ [panels (d)] between Model A and Model B is smaller than that at $z\sim 2.0$. For $z\sim 2.0$, the
galaxy sample based on Model B is shifted to lower $g$ and bluer $g-r$, but redder $u-g$ relative to that based on Model A. For $z\sim 3.0$, the galaxy
sample based on Model B is shifted to lower $g$ magnitudes but
redder $g-r$ and $u-g$ colors than those based on Model A.

\subsection{Age distribution of stellar populations}

\begin{figure}[!htp]
\begin{center}
   \includegraphics[bb=18 15 750 600,height=8.cm,width=8.cm,clip,angle=0]{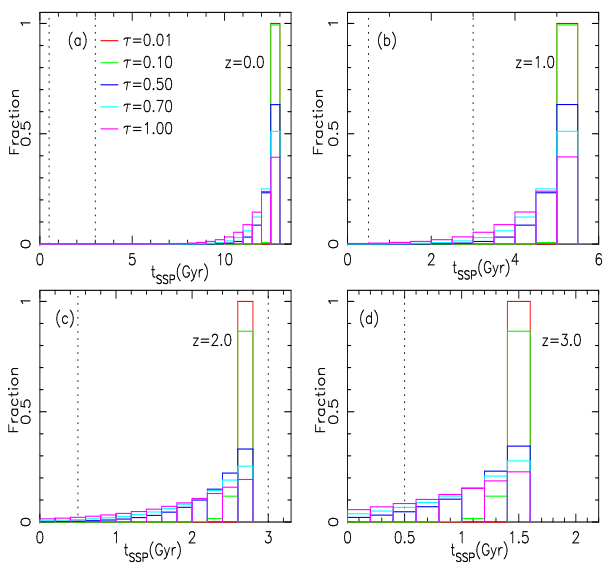}
   \caption{The age distribution of stellar populations in passive galaxies at different redshifts. Different colors of solid line represent the galaxy with different $e-$folding time scales $\tau$ ($\tau = 0.01, 0.10, 0.50, 0.70,$ and 1.0\,Gyr). The vertical dotted lines in each panel represent the lower and upper age limits for the age of 0.5 and 3.0\,Gyr, which are obvious for the binary interactions.}
   \label{fig:age-dis}
\end{center}
\end{figure}

To give detailed analyses, we display the age distribution of stellar populations in galaxies at different redshift in Fig. \ref{fig:age-dis}. Panels (a), (b), (c), and (d) correspond to $z\sim0.0, 1.0, 2.0$, and $3.0$, respectively.
The different colors show galaxies with different $e-$folding time scale $\tau$, $\tau = 0.01, 0.10, 0.50, 0.70,$ and 1.0\,Gyr. The vertical dotted lines in each panel stand for the lower and upper age limits that are important for binary interactions, 0.5 and 3.0\,Gyr. At the redshift $z\sim3.0$ we only show the lower age limit because the upper age limit is beyond the age of the Universe. We find that the ages of stellar populations are all younger than the age of the Universe at each fixed redshift.

Based on the age distribution of the stellar populations in passive galaxies in Fig. \ref{fig:age-dis}, the detailed analyses for the phenomena in Fig. \ref{fig:a-subsample} are as follows:
\begin{itemize}

\item
Using the \citet{wri06} CDM model cosmology calculator and a set of parameters, we calculate the age of the Universe at different redshifts. The ages of the Universe are about $13.7$, $5.9$, $3.0$, and $2.0$\,Gyr at redshift $z\sim 0.0, 1.0, 2.0$, and $3.0$, respectively. From the panels of Fig. \ref{fig:age-dis}, we see that the age of stellar populations tends to peak at $12.5$, $5.25$, $2.75$, and $1.5$\,Gyr for model galaxies at redshift $z\sim 0.0, 1.0, 2.0$, and $3.0$, respectively. As shown above, the effect of binary interactions is obvious for the stellar populations with age in the range of $0.5-3.0$\,Gyr. From the value of peak age and age distribution of stellar populations in model galaxies at different redshifts, we find that the ages of stellar populations in model galaxies at redshift $z\sim 2.0$ and $3.0$ are just in the age range of $0.5-3.0$\,Gyr, whereas the ages of stellar populations in the model galaxies at redshift $z\sim 0.0$ and $1.0$ are beyond this age range. This characteristic can also be seen in Fig. \ref{fig:age-dis}, where the ages of stellar populations are within the dotted lines for model galaxies at $z\sim 2.0$ and $3.0$, but outside the dotted lines for $z\sim 0.0$ and $1.0$ which verifies that the effect of binary interactions is obvious in model galaxies at redshift $z\sim 2.0$ and $3.0$, but is not obvious at redshift $z\sim 0.0$ and $1.0$. This result can explain that there exists offset for model galaxies based on Model A and Model B at redshift $z\sim 2.0$ and $3.0$, whereas this phenomenon does not exist for model galaxies at redshift $z\sim 0.0$ and $1.0$.
\item
For Fig. \ref{fig:a-subsample}, we also point out that the offset
between model galaxies based on Model A and Model B at redshift
$z\sim 3.0$ is smaller than that of model galaxies at redshift
$z\sim 2.0$. As shown in Fig. \ref{fig:age-dis}, the ages of
stellar populations in model galaxies at $z\sim 2.0$ and $3.0$
are within the age range influenced by binary interactions.
In comparison, some very young ($<0.5$\,Gyr) stellar
populations emerge in model galaxies at $z\sim 3.0$ in panel (d), which does
not appear in panel (c) at $z\sim 2.0$.
The existence of young stellar populations can also radiate the
UV-light, which is similar to the effect of binary interactions.
\citet{kav09} demonstrated that the UV-flux was highly sensitive to young stellar populations. They constructed a model in
which an old ($10.0$\,Gyr) population contributed $99.0$ percent
of the stellar mass, with a $1.0$ percent contribution from stars
that were $0.3$\,Gyr old. They found that the UV-flux of the
combined SED came purely from the young population which certifies
that the young stellar populations have large contribution to the
UV-flux, which can pollute the effect of binary interactions.

$\,\,\,\,$ In addition, the radiation in the rest-frame $F\rm_{UV}-$band can move into the observed-frame $g-$band at $z\sim 2.0$, and the observed-frame $g-$ and $r-$bands at $z\sim 3.0$. In other words, the binary interactions can affect the observed-frame $r-$band and this effect is larger than that on $g-$band for model galaxies at $z\sim 3.0$. So the $r$ magnitude becomes smaller, and makes the $g-r$ color become redder.

\end{itemize}

In general, binary interactions can influence the optical passbands (i.e., $g$-band) for passive galaxies at $z\sim2.0$. The inclusion of binary interactions in EPS models can reduce the $g-$band magnitude by $1.5$\,mag, make the $g-r$ color bluer by $1.0$\,mag, and make the $u-g$ color redder by $1.0$\,mag for the passive galaxies at redshift $z\sim2.0$.

\subsection{Dependence on the EPS models}
The validity of the results depends on the assumptions of the EPS models.
The main input parameters and distributions are (i) the common
envelope (CE) ejection $\alpha \rm_{CE}$, (ii) the coefficient $\eta$ for the Reimers wind mass loss, (iii)
the IMF of the primaries, (iv) the secondary-mass distribution,
(v) the distribution of orbital separations, and (vi) the eccentricity
distribution.

From \citet{mar11}, we know that the distribution of mass ratio ($n(q)$) is still uncertain and a matter for debate. For the EPS models, \citet{zha05} simulated the real populations by producing $2.5\times10^{7}$ binary systems by choosing three forms of $n(q)$: $n(q)=1$, $n(q)=2q$, and $M_{2}$ was uncorrelated with $M_{1}$. Moreover, they found that the discrepancy in colors caused by the choice of initial distribution of $q$ is smaller than that of the inclusion of binary interactions.

\citet{zha05} investigated the effects of  some input parameters (the CE ejection efficiency $\alpha\rm_{CE}$ and the Reimers wind mass-loss coefficient $\eta$) and input distributions (eccentricity and the initial mass of the secondaries) on the integrate colors of Model B. The results revealed that the variations in the choice of input model parameters and distributions could affect the results. For example, the circular distribution can make the colors redder than that of eccentricity distribution, while the increasing $\alpha\rm_{CE}$ can make the integrated colors bluer, and the variation of mass ratio can lead to fluctuations in the integrated colors and this fluctuation at late ages are greater than at early ages.
However, comparing the discrepancies that exist among the integrated colors, they also found that the differences between the models with and without binary interactions were greater than those caused by the variations in the choice of input parameters and distributions.

The choice of different form of IMF can possibly affect the results.
\citet{zfh12} analyzed the effect of IMF on the SFR calibration in
terms of UV luminosity for burst, S0, Sa-Sd, and Irr galaxies.
They chose the different IMF of \citet{miller79} and
\citet{salpeter55} to investigate this effect. And they found
that the effect on the UV luminosity that is caused by the variation
in the form of IMF was smaller than that between the models with and
without binary interactions for burst galaxies at all ages. So the
variation in the form of IMF can lead to small effects on the above
results. Therefore, the inclusion of binary interactions is the main
reason which causes the phenomena in Fig. \ref{fig:a-subsample}.

\section{Conclusions}
\label{sect:conclusions}
We used the galaxy template spectra built on Yunnan EPS models and $e-$folding SFR, the reddening raw of \citet{cal00}, the SDSS standard filter, and the cosmological parameters of standard dark matter model with the MC simulation to produce the passive model galaxies in the redshift range of $z=0.0$ to $3.0$, and then studied the effect of binary interactions on the predicted C-M and C-C relations with redshift. Through comparing the predicted C-M and C-C relations of galaxies, we can investigate the effect of binary interactions on these predicted relations.

For the passive galaxies, we find that the predicted C-M and C-C relations of model galaxies based on Model A and Model B have large offset at redshift $z\sim2.0$, especially for the C-C relation. These offsets are mainly produced by the inclusion of binary interaction in EPS models. At redshift $z\sim0.0$ and $1.0$, the ages of stellar populations are beyond the age range of the effect of binary interactions ($0.5-3.0$\,Gyr), therefore the effect of binary interactions is insignificant. Moreover, at redshift $z\sim3.0$, the young stellar populations ($t<0.5$\,Gyr) exist in model galaxies, which is beyond the age range of the effect of binary interactions. The existence of young stars can influence the effect of binary interactions. At redshift $z\sim2.0$, the ages of stellar populations are just in the age range of the effect of binary interactions, which is not influenced by very young stars. Therefore the effect of binary interactions are very obvious at this redshift. The binary interactions can make the $g-$band magnitude
smaller by $1.5$\,mag, the $g-r$ color bluer by $1.0$\,mag, and the $u-g$ color redder by $1.0$\,mag for the passive galaxies at this redshift. We note that the effects of different choices of input parameters and distributions for EPS models on the above results are smaller than that of inclusion of binary interactions.
Because of the widely observed optical passbands of galaxies, the stellar population properties of passive galaxies can also be affected by binary interactions. We will analyse the impact of binary interactions on the stellar population properties of observed passive galaxies in our follow-up papers.

\begin{acknowledgements}
This work is supported in part by Xinjiang Natural Science Foundation (Grant No. 2011211A104), Natural Science Foundation (Grant
Nos. 11273053, 11033008, 10821061, 2007CB815406, and 11103054), and by the Chinese Academy of Sciences under Grant No. KJCX2-YW-T24. The authors are also supported by the program of the Light in China$^,$ Western Region (LCWR) under Grants XBBS201221 and XBBS2011022.
\end{acknowledgements}

\bibliographystyle{aa}
\bibliography{zhy}

%\appendix
%\section{Appendix A}

\end{document}